\documentclass[useAMS,usenatbib]{mn2e}

\usepackage{amssymb}
                                                                                
\usepackage{times}
\usepackage{epsfig}
                                                                                
\def\be{\begin{equation}}
\def\ee{\end{equation}}

\newcommand\cm{\,{\rm cm}}
\newcommand\G{\,{\rm G}}
\newcommand\km{\,{\rm km}}
\newcommand\secinv{\,{\rm s}^{-1}}
\newcommand\invy{\,{\rm y}^{-1}}
\newcommand\erg{\,{\rm erg}}
\newcommand\cminvq{\,{\rm cm}^{-3}}

\def\d{{\rm d}}
\def\ne{n_{\rm e}}
\def\me{m_{\rm e}}
\def\Ep{E_{\rm p}}
\def\Er{E_{\rm rad}}
\def\Ek{E_{\rm K}}
\def\Ers{E_{\rm rad,52}}
\def\taut{\tau_{\rm T}}
\def\taus{\tau_{\rm s}}

\def\taucr{\tau_{\rm cr}}
\def\gamcr{\gamma_{\rm cr}}

\def\sigmat{\sigma_{\rm T}}
\def\lamb{\Lambda_{\rm B}}
\def\lams{\Lambda_{\rm T}}
\def\lamsyn{\Lambda_{\rm s}}
\def\lamcomp{\Lambda_{\rm c}}

\def\rhoe{U} 

\def\gammab{\gamma_{\rm b}}
\def\Bcr{B_{\sf QED}}

\title[Gamma-ray bursts from synchrotron self-Compton emission]
{Gamma-ray bursts from synchrotron self-Compton emission}

\author[B. E. Stern and J. Poutanen] 
{Boris~E.~Stern$^{1,2,3}$\thanks{E-mail:
stern@bes.asc.rssi.ru (BES), juri.poutanen@oulu.fi (JP)} 
and Juri~Poutanen$^{1}$\footnotemark[1]
\thanks{Corresponding Fellow, NORDITA, Copenhagen}
\\
$^{1}$Astronomy Division, P.O.Box 3000, 90014 University of Oulu,
Finland\\
$^{2}$Institute for Nuclear Research, Russian Academy of Sciences,
7a, Prospect 60-letija Oktjabrja, Moscow 117312, Russia\\
$^{3}$Astro Space Center of Lebedev Physical Institute,
Profsoyuznaya 84/32, Moscow 117997, Russia
}

\begin{document}

\date{Accepted, Received}
\pagerange{\pageref{firstpage}--\pageref{lastpage}} \pubyear{2004}
                                                                                
\maketitle

\label{firstpage}


\begin{abstract}
The emission mechanism of the gamma-ray bursts (GRBs) is still a matter 
of debates. The standard synchrotron energy spectrum of cooling electrons 
$F_{E}\propto E^{-1/2}$ is much too soft to account for the majority of the 
observed spectral slopes. 
An alternative in the form of quasi-thermal Comptonization in a 
high compactness source has difficulties in reproducing the peak of the 
observed photon distribution below a few hundred keV. 
We show here that for typical parameters expected in the
GRB ejecta the observed spectra in the 20--1000 keV BATSE energy range 
can be produced by inverse Compton scattering  
of the synchrotron radiation in a partially  self-absorbed regime. 
If the particles are continuously accelerated/heated over the life-time 
of a source rather than being instantly injected, a prominent peak 
develops in their distribution at a Lorentz factor $\gamma\sim 30-100$, 
where synchrotron and inverse-Compton losses are balanced by 
acceleration and heating due to synchrotron self-absorption. 
The synchrotron peak should be observed at 10--100 eV, while 
the self-absorbed low-energy tail with  $F_{E}\propto E^{2}$ 
can produce the prompt optical emission (like in the case of GRB~990123). 
The first Compton scattering radiation by nearly monoenergetic electrons 
can then be as hard as $F_{E}\propto E^{1}$ 
reproducing the hardness of most of the observed GRB spectra. 
The second Compton peak should be observed in the high energy gamma-ray band,
possibly being responsible for the emission detected by EGRET in GRB~941017. 
A significant electron-positron pair production reduces the available energy per 
particle, moving the spectral peaks to lower energies as the burst progresses.
The regime is very robust, operates in a broad range of parameter space and 
can explain most of the observed GRB spectra and their temporal evolution.
\end{abstract}


\section{Introduction}

Spectra of the prompt soft gamma-ray emission of gamma-ray bursts (GRBs)
are still not explained and seem mysterious despite large theoretical 
efforts devoted to this problem.
Already  in the 1980-ies it was recognized  
that synchrotron emission from the electrons injected at high energies 
produces cooling spectra $F_E\propto E^{-1/2}$ (described by  a 
photon spectral index $\alpha=-3/2$) which are much too soft 
to be consistent with that observed from GRBs \citep[e.g.][]{bus84,ie87}.
The problem became acute when \citet{pr00} showed 
that the time-resolved spectra have the mean observed $\alpha$  
close to $-1$ (i.e. $F_E\propto E^{0}$)   
and some spectra can be as hard as $F_E\propto E^1$. 
In spite of these facts, many different versions of the 
synchrotron shock models were proposed recently
\citep[see e.g.][]{t96,cd99,p99} to explain GRB spectra. 
\citet{pm00} hypothesized that inverse Compton scattering  of 
synchrotron self-absorbed radiation can be responsible for 
the observed hard BATSE (i.e. in the range 20--1000 keV) 
spectra under an assumption that electrons emit in a {\it slow} 
cooling regime, which, however, is hardly possible in the 
GRB ejecta \citep{gc99,gcl00}.

In principle, efficient cooling of electrons can be prevented by 
their reacceleration \citep{lp00}.  In synchrotron models, this 
requires the fraction of particles taking part in that process  
to be orders of  magnitudes smaller than the total number of particles 
(not to exceed the available energy) and they should always 
be the same \citep{gcl00}, conditions that are difficult to imagine. 

Problems with the relativistic synchrotron (and self-Compton) models 
gave rise to optically thick emission models such as quasi-thermal 
Comptonization \citep{zl86,t94,l97,gc99,st99}, where the energy is 
shared among many particles which are now mildly relativistic. 
If the synchrotron radiation is self-absorbed, one can achieve rather 
hard spectra with the peak at 10--50 keV in the comoving frame of the 
ejecta. For the bulk Lorentz factor $\Gamma\sim100$, this 
peak shifts, however, to an uncomfortably high energy.

In this paper we show that optically thin synchrotron self-Compton mechanism 
operating at parameters expected in the GRB ejecta can naturally 
produce very hard  spectra peaking in the BATSE energy band,
if the available energy is shared among all particles and the 
particles are continuously accelerated/heated over the life-time of a source. 
In such conditions, the electron/pair distribution develops 
a prominent peak at a Lorentz factor $\gamma\sim30-100$, 
where synchrotron and Compton losses are balanced by particle acceleration 
and heating due to synchrotron self-absorption. 
A copious pair production reduces available energy per 
particle, moving the spectral peak to lower energies as the burst progresses,
reproducing thus the  hard-to-soft evolution observed in time-resolved 
spectra \citep{ford95,rs02}.
High energy emission observed in some GRBs \citep[e.g. GRB~941017,][]{gon03},
and the prompt optical emission observed in GRB~990123 \citep{a99} also can be 
explained in this model simultaneously.

\section{Main parameters}
\label{sec:param} 

Let us consider the ejecta moving with Lorentz factor $\Gamma$
at the distance $R$ from the source. The main parameters 
determining radiation physics are the (comoving) energy dissipation rate, 
magnetic field strength $B$, size of the emission region $R'$, 
and the number of particles described by the Thomson optical depth 
$\taut=\ne R' \sigmat$. It is suitable to describe
the available energy  by the comoving compactness
\be\label{eq:defcomp}
\Lambda = {\rhoe \over \me c^2} R' \sigmat , 
\ee
where $\rhoe$ is the comoving density of a relevant kind of energy.  Formally,
$\Lambda$ is just the optical depth $\taut$ of pairs if we would spend all
available energy for their mass. In reality, most energy goes to radiation 
and the resulting $\taut \ll \Lambda$ (e.g., $\taut \sim 10 - 20$ 
at $\Lambda = 10^3$ and $\taut \sim 1$ at $\Lambda = 30$). 
The role of magnetic fields can be described by 
the magnetic compactness $\lamb$ given by Eq.~(\ref{eq:defcomp}) with 
$\rhoe=B^2/(8\pi)$. The ratio $\lamb/\Lambda$ is model dependent.
We assume that the magnetic field is below equipartition, i.e.
$\lamb\lesssim \Lambda$. Even in magnetically dominating models, 
one does not expect necessarily $\lamb\gg\Lambda$, since reconnection
of magnetic field  providing the energy dissipation can reduce its 
strength {\it within} the emission region to $\lamb\sim\Lambda$.

\begin{figure}
\centerline{\epsfig{file=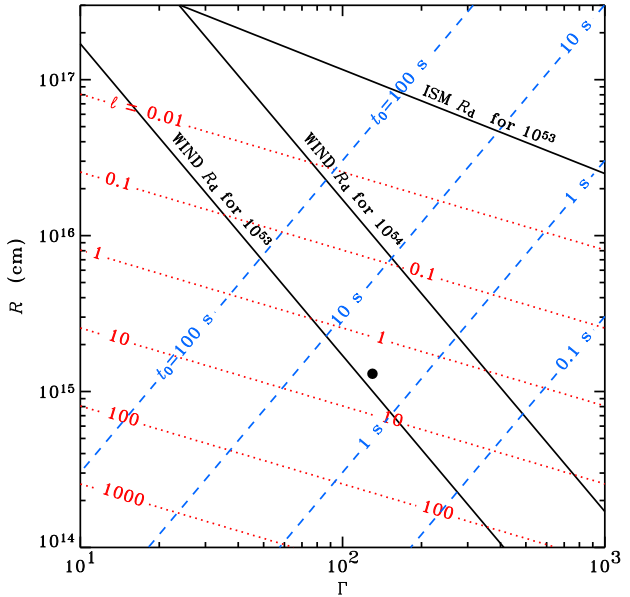,width=8.cm}}
\caption{
The map of $\Gamma-R$ parameter space. Dotted lines show constant 
compactness $\ell$ levels for energy release $\Er=10^{52} \erg$ in the causally 
connected region given by Eq.~(\ref{eq:comp}). 
Dashed lines show the time-scale in the observer frame defined as  
$t_0 = R/(c\Gamma^2)$. 
The deceleration limit $R_{\rm d}$ for different $\Ek$ 
is given for the wind with the mass loss rate 
$10^{-5} M_\odot \invy$ and velocity of $10^3\km\secinv$ as well as 
for the interstellar medium of $n_{\rm ISM}=1$ (see Eq.~[\ref{eq:raddec}]).
The circle corresponds to the model run simulated in the paper.
}   
\label{fig:map}
\end{figure}

It is evident that $R'$ should not exceed the size of the causally
connected region, i.e. $R/\Gamma$ in both transversal and
radial direction (the latter in the observer's frame is $R/\Gamma^2$).
We assume $R' = R/\Gamma$ and measure the comoving time $t'$ in units of 
the light crossing time of the region $R'/c=R/(c\Gamma)$ which corresponds to the 
observer time $R/(c\Gamma^2)$. The dissipation compactness (corresponding to the energy 
dissipation {\it rate}) is then $\ell=\d\Lambda/\d t'$. 
(For a constant dissipation rate during $t'=1$ we get $\ell=\Lambda$.) 
It can be estimated from the observed isotropic energy release $\Er$ 
assuming that the energy was dissipated  homogeneously in a causally 
connected region  
\be\label{eq:comp}
\ell = {\Er \sigmat \over \me c^2 \Gamma 4 \pi R^2} = 
7\ \Ers \Gamma_2^{-1} R_{15}^{-2} .
\ee
(Here we adopt notation $Q=10^x Q_x$ in cgs units if not mentioned 
otherwise.) 
Figure~\ref{fig:map} demonstrates the levels of the  compactness and 
the observed time-scale on a  $R - \Gamma$ plane. 
The observed emission episode at the first glance should be 
a single pulse of time-scale $R/(c\Gamma^2)$. Actually, most of GRBs 
have a complex  time structure. We can then prescribe this episode to
a single GRB pulse and associate $\Er$ with its energy fluence, or 
to admit that the energy can be released by compact flares 
within the causally connected region. In that case, $\Er$ should be referred to 
the fluence of a complex emission episode and Eq.~(\ref{eq:comp}) then 
gives an average compactness in the region while the local values of $\ell$ 
can be substantially higher.

What is the typical size $R$ where the energy dissipation takes place?
For radii smaller than $\lesssim10^{14}\cm$, the compactness is large 
$\ell\gtrsim300$ (see Fig.~\ref{fig:map}), pair production is extremely 
efficient, and the mean energy available per particle is rather small. 
The main emission mechanism is then multiple Compton scattering.
The spectra expected from this regime can be hard \citep{gc99,st99}, 
but peaked at too high energies (10--50 keV in the comoving frame).
At $R\gtrsim 10^{15}\cm$, the compactness and optical depth are smaller,
the mean energy available per particle larger and the energy is radiated 
away by optically thin or synchrotron or synchrotron self-Compton emission.

The dissipation radius is also limited from above by 
deceleration of ejecta in the external environment 
\begin{equation} \label{eq:raddec}
R_{\rm d} \sim \left\{ 
\begin{array}{ll} 
1.7\times 10^{16}\ E_{\rm K,54}\ w_{3} 
\dot{M}_{-5}^{-1} \Gamma_{2}^{-2}  \cm,  & \mbox{for\ wind,} \\
2.5\times 10^{17}\ E_{\rm K,54}^{1/3}\  
n_{\rm ISM} ^{-1/3} \Gamma_{2}^{-2/3}  \cm,  & \mbox{for\ ISM,} 
\end{array} \right.
\end{equation} 
where the ejecta sweeps up mass $\sim M_{\rm ej}/\Gamma$ 
(see solid lines in Fig.~\ref{fig:map}). Here $E_{\rm K}=\Gamma M_{\rm ej}c^2$ 
is the isotropic kinetic energy of the ejecta, $n_{\rm ISM} \cminvq$ is particle 
concentration in the interstellar medium,
$\dot M_{-5}$ is the mass loss rate by the progenitor star in units
$10^{-5} M_\odot \invy$, and $w_{3}$ is the wind velocity in units $1000\km\secinv$.
We notice that the deceleration limit in the wind case 
restricts the low compactness ($\ell \lesssim 10$) 
regime to a relatively narrow region of parameters. 
If the kinetic energy is $\lesssim 10^{53}\erg$, or the environment  is denser, 
then the low-compactness regime is hardly possible at all. 
Thus we consider the typical dissipation radius $R=10^{15}\cm$. 
The observed pulse duration of about 1 s is consistent 
with the fact that GRBs have  very little power at time-scales below 1 s
\citep{bss00}. 

The optical depth $\taut$ in the emission region is bounded from above by
the opacity of the ejecta 
\be
\tau_{\rm ej} = 0.3\  E_{\rm K,54}\ R_{15}^{-2}\ \Gamma_{2}^{-1} 
\ee
(assuming  matter dominated ejecta and that 
the matter is concentrated within a causally connected shell). 
For external shocks,  $\taut$ cannot be smaller  than that determined by 
collected material
\be \label{eq:tauext}
\tau_{\rm ext} = \left\{ \begin{array}{ll}  
2\times 10^{-4}\ \dot{M}_{-5}\ R_{15}^{-1}\ w_{3}^{-1} ,& \mbox{for\ wind,} \\
 2\times 10^{-8}\  R_{17}\ n_{\rm ISM}, & \mbox{for\ ISM.} 
\end{array} \right.
\ee

What is the physical mechanism of the energy dissipation? 
In the external shock models \citep{rm92,p99}, 
the dissipation occurs at $R\sim R_{\rm d}$ which can be about 
$10^{15}\cm$ for the wind environment 
if $\Ek\sim 10^{53}\erg$ and $\Gamma\sim200$ 
(see Fig.~\ref{fig:map}).
The impulsive first-order Fermi acceleration would 
operate in a fast cooling regime \citep[see][for a specific 
version of this scenario]{st03} which, as discussed in the Introduction, 
contradicts the data. 
An alternative version of shock energy dissipation 
is particle heating by plasma instabilities behind the shock front 
\citep[see e.g.][]{fre03}. We follow this route here.
It can also operate in internal or `refreshed' shocks which could be 
produced  by collisions of fresh ejecta with previously ejected, 
partially decelerated material.
In the Poynting flux dominated models \citep{usov94,lb03}, the magnetic 
field energy can be dissipated at the required distances.

\section{Physical model}

\subsection{Model setup}

As we have seen  a number of models can satisfy our requirements.
Thus we do not specify the exact model for the energy dissipation, but 
consider a toy model where energy is injected to the emission region 
with the constant rate during comoving time $R'/c$.
We adopt a slab geometry of the emission region which, in 
a zero approximation, is consistent with both
internal and external shock scenarios. Indeed, we can expect that the main 
energy release takes place behind the shock front in a layer which is 
thin relative to the size of the causally connected region.
This geometry can also be a satisfactory approximation for the 
magnetic reconnection scenario. Indeed, the magnetic field is
probably predominantly transversal  and the reconnection plane 
is again perpendicular to the direction of propagation. 

We neglect the curvature of the shock front and the bulk velocity
gradient thus reducing the problem to a static slab. In this approach we 
omit a number of effects associated with relativistic expansion of the 
emitting shell. These effects are important for the description of the 
time evolution, however they are not critical for understanding 
general spectral properties.
The thickness of the emitting slab, $\Delta<1$ (in units of $R'$),
is unknown since it depends on the relaxation process behind the shock 
front and is probably smaller in the case of the reconnection scenario. 
We take $\Delta=0.1$, but the results are not very sensitive to its value.
 
The energy release is uniform over the slab volume and we assume 
that the energy is injected in a form of acceleration of electrons and 
pairs which  obtain equal amount of energy per unit time. 
The optical depth can increase due to pair production. 
We treat the magnetic field geometry as chaotic, therefore all pairs are isotropic.

The model is fully described by four parameters: 
(i) the initial Thomson optical depth across the slab, 
$\tau_0=\ne\sigmat \Delta\ R'$; 
(ii) comoving size $R'$; (iii) dissipation compactness $\ell$; 
and (iv) magnetic compactness $\lamb$.

\subsection{Radiative processes}

Let us first consider how particles (electron and positrons) 
of Lorentz factor $\gamma$
are heated and how do they cool. The energy gain rate of a 
particle is simply given by the heating rate $\propto \ell$ divided by the 
number of particles which is proportional to the total Thomson optical depth 
(including pairs) across the slab $\taut$. Particles cool
by emitting synchrotron radiation and by scattering this radiation
(self-Compton mechanism). The energy balance equation 
can be written as:
\be \label{eq:enebal}
\frac{\d \gamma}{\d t'}= \frac{\ell}{\taut} - 
\frac{4}{3}(\eta\lamb + \lams)\gamma^2 .
\ee
Here $\eta<1$ accounts for the reduced synchrotron cooling 
due to synchrotron self-absorption, and 
$\lams$ is the compactness corresponding to the energy density
of soft photons in the Thomson regime (with energy 
$\epsilon\equiv E/ \me c^2 \lesssim 1/\gamma$). 
The typical cooling time is then 
$t_{\rm cool}  \sim 1/[(\lams + \eta\lamb) \gamma ]$, 
which, for the GRB conditions, 
is orders of magnitude smaller than the light-crossing time.

The balance between heating and cooling is achieved at 
\be \label{eq:gampeak}
\gammab\approx \sqrt{\ell/(\lams+\eta\lamb)} \taut^{-1/2} ,
\ee
where $\lams$ and $\eta$ also depend on $\gammab$.
Particles with $\gamma>\gammab$ lose energy faster than they gain it, 
while at $\gamma<\gammab$ the situation is opposite. 
As a result, a very narrow electron distribution peaked at $\gammab$ develops.
This allows us to adopt the approximation that all particles have the same 
Lorentz factor $\gamma=\gammab$. 
The radiation compactness $\lams$ can be expressed as a sum
of the synchrotron $\lamsyn=y \eta\lamb$ 
and first Compton scattering $\lamcomp=y \eta\lamsyn$ 
energy densities (further scattering orders are in the Klein-Nishina limit).
In the adopted approximation, the Compton parameter is just $y=\xi\taut \gamma^2$,
where the geometrical factor $\xi\sim 1$ for a spherical source and 
$\xi\sim (2/3) \ln (3/2\Delta)\sim 1.8$ for a slab with $\Delta=0.1$.
Thus Eq.~(\ref{eq:gampeak}) is reduced to
\be \label{eq:yyylll}
y(1+y+y^2)\approx \xi \ell /(\eta\lamb).
\ee
When synchrotron self-absorption is negligible, $\eta=1$, 
we find the solution $y_0\approx (\xi\ell/\lamb)^{1/3}$ 
(or $y_0\approx (\xi\ell/\lamb)^{1/2}$, if the second Compton 
scattering operates close to the Klein-Nishina limit). 
At small  $\eta$, Compton parameter increases.

The importance of  self-absorption depends crucially on $\gamma$.
The  optical depth at frequencies 
below the synchrotron emission peak is \citep[eq. 2.18a in][]{gs91}
\be\label{eq:taus}
\taus = 15 \taut/(b\gamma^5 x^{5/3}) , \qquad x=\epsilon/(3\gamma^2 b),
\ee
where $b=B/\Bcr$ and $\Bcr=4.4\times10^{13}\G$.
Thus the emission will be significantly reduced if the self-absorption frequency 
(where $\taus=1$) is above the emission peak ($x\gtrsim1$). This happens
at $\gamma < \gamcr = 50 (\tau_{\rm T, -3}/B_3)^{1/5}$. The same condition 
for $\taut$ expressed via Compton parameter is 
$\taut > \taucr(y)=0.5\times 10^{-3} (y/\xi)^{5/7}  B_3^{2/7}$. 
 
We now can predict the temporal evolution of the radiation spectrum. 
The optical depth starts growing after about two light-crossing times $2\Delta$
required to produce high energy photons. 
If $\taut<\taucr(y_0)$, the synchrotron is not absorbed and $y=y_0$,
$\gamma^2=y_0/(\xi\taut)$, 
and the synchrotron peak energy decreases with optical depth as 
 $\epsilon_{\rm s}\sim 3\gamma^2 b \sim 
2\times 10^{-7}(y_0/\xi)^{2/7} B_3^{5/7}\taucr/\taut$.  
The first Compton peak evolves even faster $\epsilon_{\rm c1}\sim (4/3) 
\gamma^2 \epsilon_{\rm s} \sim 4\times 10^{-4}(y_0/\xi)^{4/7} B_3^{3/7}
(\taucr/\taut)^2$. 
When $\taut$ grows above $\taucr$, $\eta$ decreases  because of self-absorption, 
$y$ correspondingly increases (Eq.~\ref{eq:yyylll}), the synchrotron peak 
becomes more stable, and the resulting electron energy and the first Compton 
peak start to evolve slower. 
The second Compton peak is produced in the Klein-Nishina limit at small $\taut$, 
and evolves slowly $\epsilon_{\rm c2}\sim \gamma \propto 1/\sqrt{\taut}$, while 
at larger $\taut$ the evolution speeds up. 
We now check there predictions by numerical simulations. 

\subsection{Simulations}
\label{sec:results}

The simulations were performed using a Large Particle Monte-Carlo code 
(LPMC) developed by \citet{st85} and \citet{st95}. 
It treats Compton scattering, 
photon-photon pair production and  pair annihilation, 
synchrotron radiation and synchrotron self-absorption. 
The exact cross-section are used for the first three processes,
while the cross-sections in the relativistic approximation 
are used for the synchrotron process \citep{gs91}.
The electron/pair and photon distributions are computed self-consistently. 
The code is essentially nonlinear: the simulated particles constitute
at the same time a target medium for other particles. 
The geometry of the emission region is a pill-box of radius $R'$
and thickness $\Delta=0.1$. Output photons are recorded when they cross the 
surfaces $z = \pm(\Delta /2 + 0.1 )R'$.

\begin{figure}
\centerline{\epsfig{file=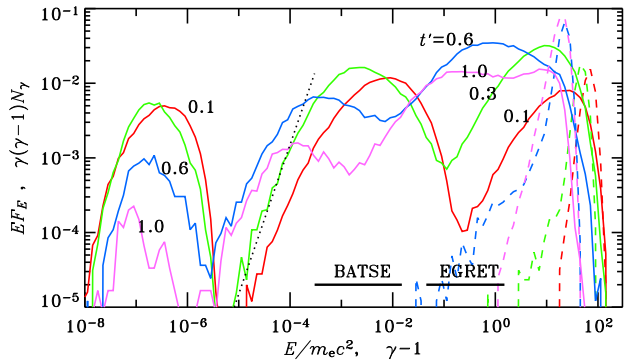,width=8.5cm}}
\caption{
Instantaneous  (comoving frame) photon spectra 
within the slab (i.e. the source functions) and corresponding electron 
distributions. 
Solid curves show the photon spectra $EF_E$ (in arbitrary units) 
at times of $0.1, 0.3$,  $0.6$, and $1$ (in units $R'/c$). The spectra consist
of a low-energy (partially self-absorbed) synchrotron bump and two 
Compton scattering orders. Further scatterings are suppressed due to the 
Klein-Nishina effect.  The electron energy distribution function 
$\gamma(\gamma-1) \d N/\d\gamma$ for the same 
time intervals is shown by dashed  curves (the peak evolves towards lower energies).
Parameters of simulations: $R'=10^{13}\cm$, $\tau_0=6\times 10^{-4}$, $\ell=3$, 
$\lamb=0.3$. Dotted line shows the hardest 
possible $F_E\propto E^1$ power-law reachable at the low-energy slope of the first 
Compton bump. The BATSE (20-1000 keV) and EGRET (3-100 MeV) bands 
redshifted to a comoving frame by $(1+z)/2\Gamma$ 
(where $z\sim 1$ is the cosmological redshift and $\Gamma=130$) 
are shown by horizontal bars. 
}
\label{fig:spectra}
\end{figure}

As an example, we take typical parameters described in 
\S~\ref{sec:param}. The comoving radius $R'=10^{13}\cm$,
the initial Thomson optical depth $\tau_0=6\times 10^{-4}$ 
(close to the critical $\taucr$, but higher than $\tau_{\rm ext}$),  
compactness $\ell=3$ (corresponding to $\Gamma\approx 130$ for 
$\Ers=1$, see Eq.~\ref{eq:comp}), 
and the magnetic compactness $\lamb=0.3$ (corresponding to the 
comoving magnetic field $B'\sim 1000\G$, three times below the equipartition).
The evolution of broad-band spectra and the electron distribution 
are shown in Fig.~\ref{fig:spectra}.

\begin{figure}
\centerline{\epsfig{file=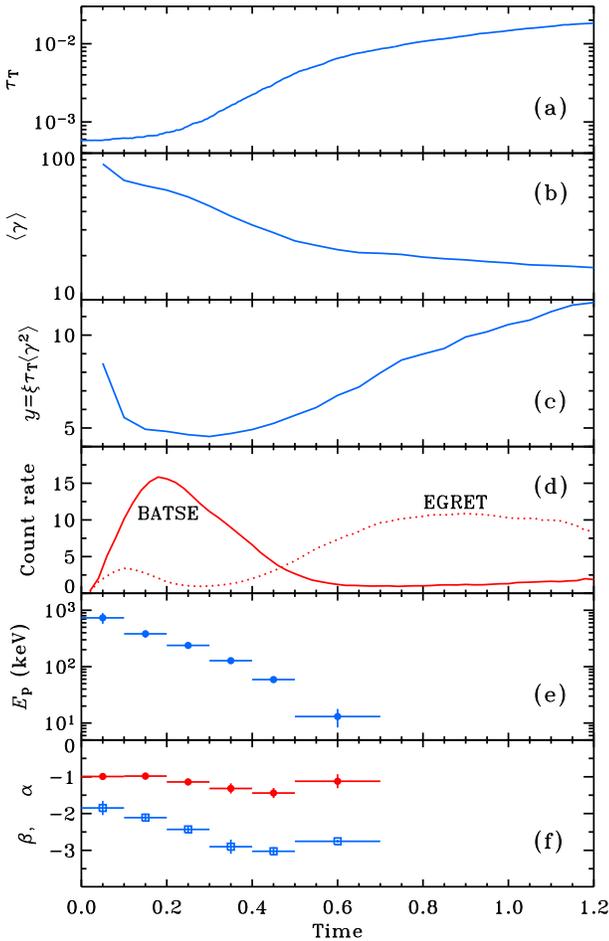,width=8.cm}}
\caption{
Evolution  (a) of the Thomson optical depth due to pair production, 
(b) of the mean electron Lorentz factor $\langle \gamma \rangle$, 
and (c) of the Compton $y$-parameter. 
(d) Photon flux in the BATSE band (solid curve) and the EGRET band 
(dotted curve) in arbitrary units. 
Evolution of spectral parameters (e) $\Ep$, 
(f) $\alpha$ (circles), and $\beta$ (squares).
}
\label{fig:all}
\end{figure}

At the start of simulations, $\taut=\tau_0<\taucr$ and the electron 
heating and cooling are balanced at high $\gamma$ (see dashed curves). 
Partially self-absorbed synchrotron radiation (solid curves, lower energy bump) 
peaks at $\epsilon_{\rm s}\sim 3\times 10^{-7}$ in the comoving frame and 
has a hard low-energy tail $F_E\propto E^2$ \citep{gs91}.
The optical depth grows nearly linearly with time 
due to electron-positron pair production 
(see Fig.~\ref{fig:all}a) and the mean particle energy decreases as
$\langle\gamma\rangle \propto 1/\sqrt{\tau}$ (Fig.~\ref{fig:all}b).  
The Compton parameter  $y=\xi\taut\langle\gamma^2\rangle$ 
computed from the pair distribution (see Fig.~\ref{fig:all}c) 
is about 8 in the beginning, reaches minimum at $t'=0.3$ and 
grows to more than 10 by the end of energy injection.
The `observed' $y$-parameter, i.e. the 
ratio of luminosities in the first Compton bump to the 
synchrotron component grows monotonically, however, from 3 at $t'=0.1$ 
to 16 at $t'=1$. This discrepancy is caused by the 
non-stationarity of the problem -- it takes too long time to build up 
the spectrum and to reach a steady-state.

The first Compton peak $\propto\langle\gamma^2\rangle$
moves to softer energies and crosses the `BATSE window'
(note that due to self-absorption, synchrotron peak energy is very stable). 
This spectral evolution is  consistent with the observed in time-resolved 
pulses \citep[e.g.][]{ford95,rs02}. 
The photon flux in the BATSE band  (Fig.~\ref{fig:all}d) shows 
a `fast rise -- exponential decay' behaviour often seen in GRBs. 
It  decays much before the energy supply terminates, but 
has a long flat part.  
While the second Compton peak at $\sim 10$--$100$ MeV, on the contrary, 
rises later and decays  on longer time-scale.

We fitted the  model photon spectrum $N_E=F_E/E$ in the BATSE window 
by a phenomenological `GRB function' consisting of a power-law with an exponential 
cutoff, $N_E\propto E^{\alpha}\exp(-E(2+\alpha)/\Ep)$, merging to a high-energy 
power-law $\propto E^{\beta}$  \citep{b93}.
The evolution of the fitted (observed) peak energy $\Ep$ of the $EF_E$ spectrum  
and the spectral slopes is shown in Fig.~\ref{fig:all}e,f.
One sees softening of the spectrum as the burst progresses. 
The fitted $\alpha$ is close to $-1$, the most probable value in the distribution 
of time-resolved spectra \citep{pr00}. 
The results of spectral fitting depend  somewhat on the used energy interval: 
$\alpha$ becomes larger (spectrum hardens) in a wider interval  and $\beta$ 
is softer.  
Often the data at lower energies (with better statistics) dominate 
the fitting procedure,  then the fitted $\alpha$ can be much harder. 
A correlation between $\alpha$ and $\Ep$ is also expected. 
The hardest possible spectrum, $\alpha=0$,  corresponds to Compton scattering 
by isotropic monoenergetic electrons \citep{bg70,rl79}.

\section{Summary}

The radiative processes that are responsible for particle 
cooling depend  on the optical depth in the emission region. 
If $\taut$ is very small, the mean $\gamma$ is large
and the BATSE photons would be produced by synchrotron radiation. 
At $\taut\sim 1$, quasi-thermal Comptonization operates. 
At intermediate $\taut \sim3\times 10^{-4}$--$10^{-2}$, 
the electrons/pairs have $\gamma\sim 30-100$  and 
the first Compton peak is produced in the BATSE band.
Such optical depth seems natural for the external shock 
in the typical Wolf-Rayet progenitor wind as well as 
for the emission within the ejecta (e.g. due to magnetic reconnection 
or collisions of the fresh ejecta with already decelerated material).
Efficient pair production at intermediate compactnesses, $\ell=0.1$--$10$, 
also can be responsible for the required $\taut$.
  
Synchrotron self-Compton emission from continuously heated,
nearly monoenergetic electrons can explain many observed features of GRBs. 
The synchrotron spectrum peaking at $\epsilon \sim  
3 \times 10^{-7}$ in the comoving frame will be blueshifted 
to the extreme UV region. The self-absorbed low-energy tail is hard,
$F_{E}\propto E^2$, and can explain the prompt optical radiation 
detected from GRB~990123 \citep{a99}.
The energy  of the first Compton peak is expected to decrease as the burst 
progresses since pair production reduces the mean energy available per particle.  
In the case of a larger $\tau_0$ and/or a higher compactness and/or 
a smaller $\Gamma$, the first Compton component peaks in X-rays and 
possibly can be identified with the observed X-ray flashes 
\citep[e.g.][]{hei03}.

The GRB spectral hardness distribution \citep{pr00} can also
be reproduced (maybe except its hardest events, see \citealt*{gcg03}). 
Since the incident synchrotron spectrum is hard, 
the low energy slope of the scattered radiation, $F_E\propto E^1$,
is determined by kinematics of single Compton scattering 
by monoenergetic electrons. 


The second inverse Compton peak observed at 10 MeV -- 10 GeV 
is delayed relative to the soft gamma-ray emission and 
lasts longer. In spite of large $y$-parameter, it does not 
necessarily dominate the total energy output because of the 
Klein-Nishina effect.
The rather hard ($\alpha \sim -1$) spectrum at $\sim 10$ MeV can 
match observations of GRB~941017 \citep{gon03} and its observed slow 
evolution. 
If the mean particle energy decreases rapidly, this 
component can possibly produce even  the second distinct pulse 
in the BATSE range.

Summarizing, the proposed model can explain a 
large fraction of GRB spectra and their time evolution. 
It also reproduces the high energy $\sim 10-100$ MeV emission detected by 
EGRET in some bursts and the prompt optical emission. 
Two latter phenomena  are natural within this model
and do not require additional assumptions or separate emission regions.

\section*{Acknowledgments}

We thank Marek Sikora for useful  discussions and suggestions.
This research has been supported by the RFBR grant 04-02-16987, 
Academy of Finland, Jenny and Antti Wihuri Foundation, 
Vilho, Yrj\"o and Kalle V\"ais\"al\"a Foundation, and the NORDITA 
Nordic project in High Energy Astrophysics.

\label{lastpage}

\end{document}